\documentclass[11pt,twoside]{article}
\usepackage{macro-fsut-eng}
\usepackage{graphicx}

\usepackage[T1]{fontenc} 

\usepackage{latexsym}
\usepackage{verbatim}
\usepackage{mathrsfs}
\usepackage{amsmath}
\usepackage{amssymb}
\usepackage{epsfig}

\newcommand{\postscript}[2]{\setlength{\epsfxsize}{#2\hsize}
   \centerline{\epsfbox{#1}}}

\def\deg{\ifmmode{^{\circ}}\else ${^{\circ}}$\fi}
\def\bi{\begin{itemize}}
\def\ei{\end{itemize}}
\def\bfl{\begin{flushleft}}
\def\efl{\end{flushleft}}
\def\ed{\end{document}}

\def\cf#1{\ifmmode{\cal #1}\else${\cal #1}$\fi}
\def\ra{\rightarrow}
\def\be{\begin{equation}}
\def\ee{\end{equation}}

\def\beas{\begin{eqnarray*}}
\def\eeas{\end{eqnarray*}}
\def\bea{\begin{eqnarray}}
\def\eea{\end{eqnarray}}

\newcommand{\lb}{\left(}
\newcommand{\rb}{\right)}

\newcommand{\lsb}{\left[}
\newcommand{\rsb}{\right]}

\begin{document}

\vskip 1.0cm
\markboth{L. A. Anchordoqui}{Constraints on cosmological parameters
  from Planck and BICEP2 data}
\pagestyle{myheadings}

\vspace*{0.5cm}
\title{Constraints on cosmological parameters from Planck and BICEP2 data}

\author{Luis A. Anchordoqui}
\affil{Department of Physics and Astronomy,\\
Lehman College, City University of New York, Bronx NY 10468, USA}

\begin{abstract}
  We show that the tension introduced by the detection of large
  amplitude gravitational wave power by the BICEP2 experiment with
  temperature anisotropy measurements by the Planck mission is
  alleviated in models where extra light species contribute to the
  effective number of relativistic degrees of freedom. We also show
  that inflationary models based on $S$-dual potentials are in
  agreement with Planck and \mbox{BICEP2 data.}\end{abstract}

\section{Fitting $\Lambda$CDM + $r$ to Planck and BICEP2 data}

Measurements of the cosmic microwave background (CMB) and large scale
structure (LSS) indicate that we live in a spatially-flat,
accelerating, infinite universe composed of 4\% of baryons ($b$), 26\%
of (cold) dark matter (CDM), and 70\% of dark energy
($\Lambda$). These observations also reveal that the universe has tiny
ripples of adiabatic, scale-invariant, Gaussian density perturbations.
The favored $\Lambda$CDM model implicitly includes the hypothesis of a
very early period in which the scale factor of the universe expands
exponentially: $a \propto e^{Ht}$, where $H = \dot a/a$ is the Hubble
parameter (see e.g. Baumann 2009). If the interval of exponential
expansion satisfies $\Delta t > N/H$, with $N$ above about 50 to 60,
a small casually connected region can grow sufficiently to accommodate
the observed homogeneity and isotropy, to dilute any overdensity of
magnetic monopoles, and to flatten the spatial hyper-surfaces (i.e.,
$\Omega \equiv \frac{8\pi \rho}{3 M_{\rm Pl} H^2} \to 1$, where $M_{\rm PL}
= G^{-1/2}$ is the Planck mass and $\rho$ the energy density;
throughout we use natural units, $c = \hbar = 1$). Quantum
fluctuations during this inflationary period can explain the observed
cosmological perturbations.

Fluctuations are created quantum mechanically on subhorizon scales 
with a spectrum of wavenumbers $k$. (A mode $k$ is called superhorizon
when $k < aH$ and subhorizon when $k > aH$.)  While comoving scales,
$k^{-1}$, remain constant the comoving Hubble radius, $(aH)^{-1}$,
shrinks quasi-exponentially during inflation (driving the universe
toward flatness) and the perturbations exit the horizon. Causal
physics cannot act on superhorizon perturbations and they freeze until
horizon re-entry at late times. A mode exiting the horizon can then be
described by a classical probability distribution with variance given
by the power spectrum ${\cal P}_\chi (k)$.  After horizon re-entry the
fluctuations evolve into anisotropies in the CMB and perturbations in
the LSS. The scale-dependence of the power spectrum is defined by the
scalar spectral index, $n_s -1 \equiv d \ln {\cal P}_\chi/d \ln k$,
and its running $\alpha_s \equiv dn_s/d\ln k$.  The power spectrum is
often approximated by a power law form: $ {\cal P} (k) = A_s (k_*) \lb
k /k_* \rb^{n_s -1 + \frac 1 2 \alpha_s \ln \lb \frac k {k_*} \rb +
  \cdots}, $ where $k_*$ is an arbitrary reference that typifies
scales probed by the CMB.

The Planck temperature spectrum at high multipoles ($l \gtrsim 40$)
describes the standard spatially-flat $\Lambda$CDM 6-parameter model
$\{\Omega_b h^2,\, \Omega_{\rm CDB} h^2,\, \Theta_s,\, \tau,\, n_s,\,
A_s\}$ with high precision: {\it (i)}~baryon density, $\Omega_b =
0.02207 \pm 0.00033$; {\it (ii)}~CDM density, $\Omega_{\rm CDM} h^2 =
0.1196 \pm 0.0031$; {\it (iii)}~angular size of the sound horizon at
recombination, $\Theta_s = (1.04132 \pm 0.00068) \times 10^{-2}$; {\it
  (iv)}~Thomson scattering optical depth due to reionization, $\tau =
0.097 \pm 0.038$; {\it (v)}~scalar spectral index, $n_s = 0.9616 \pm
0.0094$; {\it (vi)}~power spectrum amplitude of adiabatic scalar
perturbations, $\ln (10^{10}\, A_s) = 3.103 \pm 0.072$ (Ade et
al. 2013a). Planck data also constrain the Hubble constant $h = 0.674
\pm 0.012$ and $\Omega_\Lambda = 0.686 \pm 0.020$. (Herein we adopt
the usual convention of writing the Hubble constant at the present day
as $H_0 = 100 \ h~{\rm km} \ {\rm s}^{-1} \ {\rm Mpc}^{-1}$.) Note,
however, that the data only measure accurately the acoustic scale, and
the relation to underlying expansion parameters (e.g., via the
angular-diameter distance) depends on the assumed cosmology, including
the shape of the primordial fluctuation spectrum. Even small changes
in model assumptions can change $h$ noticeably. Unexpectedly, the
$H_0$ inference from Planck data deviates by more than $2\sigma$ from
the previous result from the maser-cepheid-supernovae distance ladder
$h = 0.738 \pm 0.024$ (Riess et al. 2011).  The impact of the Planck
$h$ estimate is particularly important in the determination of the
number of ``equivalent'' light neutrino species: $N_{\rm eff}$
(Steigman et al. 1977). Combining observations of the CMB with data
from baryon acoustic oscillations (BAO), the Planck Collaboration
reported $N_{\rm eff} = 3.30 \pm 0.27$ (Ade et al. 2013b). However, if
the value of $h$ is not allowed to float in the fit, but instead is
frozen to the value determined from the maser-cepheid-supernovae
distance ladder the Planck CMB data then gives $N_{\rm eff} = 3.62 \pm
0.25$, which suggests new neutrino-like physics (at around the
$2.3\sigma$ level).

Inflation also produces fluctuations in the tensor part of the spatial
metric. The gravity-wave fluctuations are also frozen on super-horizon
scales and their $B$-mode power spectrum, $ {\cal P}_h = A_t \lb \frac
k {k_*} \rb^{n_t + \frac 1 2 \alpha_t \ln \lb \frac k {k_*} \rb +
  \cdots},$ can be imprinted in the CMB temperature and
polarization. We define the tensor-to-scalar
amplitude ratio $r= A_t/A_s$ as the free parameter for the
$\Lambda$CDM + r model.

\begin{figure}[tbp]
\begin{minipage}[t]{0.32\textwidth}
\postscript{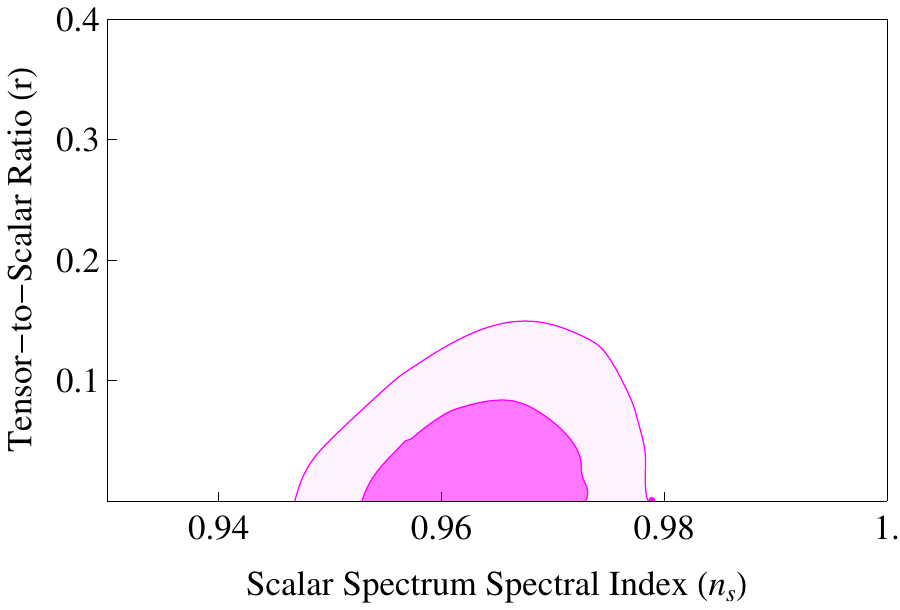}{0.99}
\end{minipage}
\hfill
\begin{minipage}[t]{0.32\textwidth}
\postscript{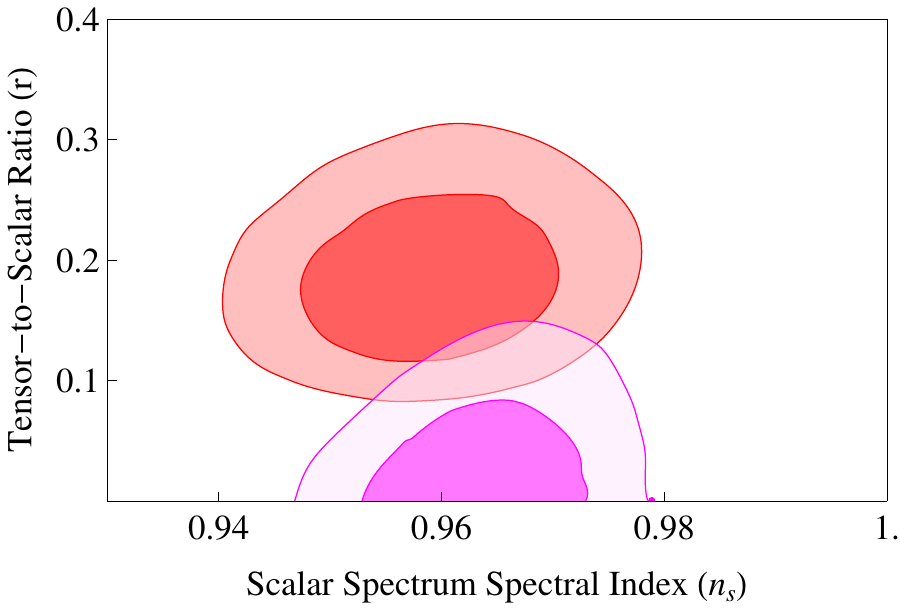}{0.99}
\end{minipage}
\hfill
\begin{minipage}[t]{0.32\textwidth}
\postscript{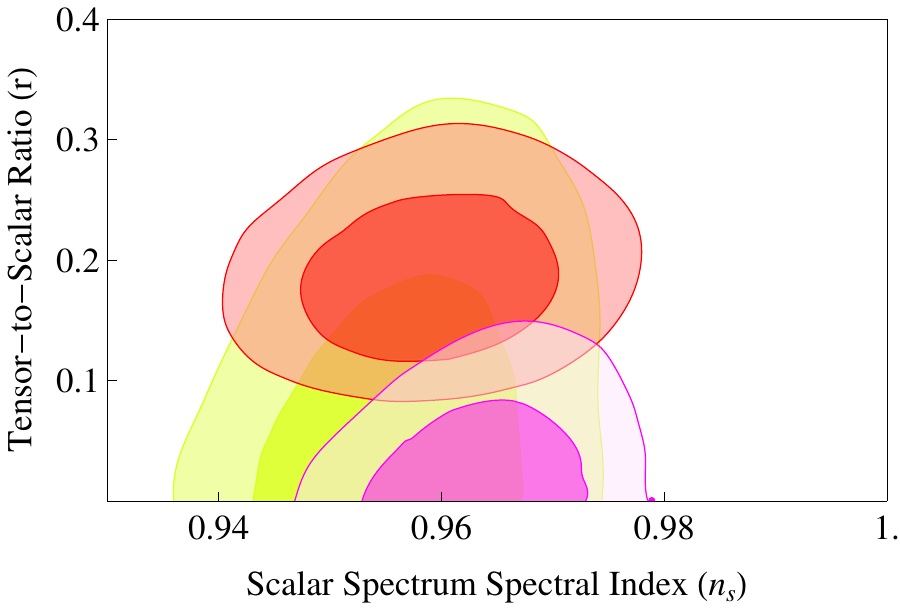}{0.99}
\end{minipage}
\caption{Marginalized joint 68\% CL and 95\% CL regions for ($r,n_s$)
  using Planck + WMAP + BAO data without a running spectral index
  (left), BICEP2 data with $\alpha_s \neq 0$ (middle), and Planck +
  WMAP + BAO data with $\alpha_s \neq 0$ (right).}
\label{fig:uno}
\end{figure}
\begin{figure}[tbp]
\begin{minipage}[t]{0.32\textwidth}
\postscript{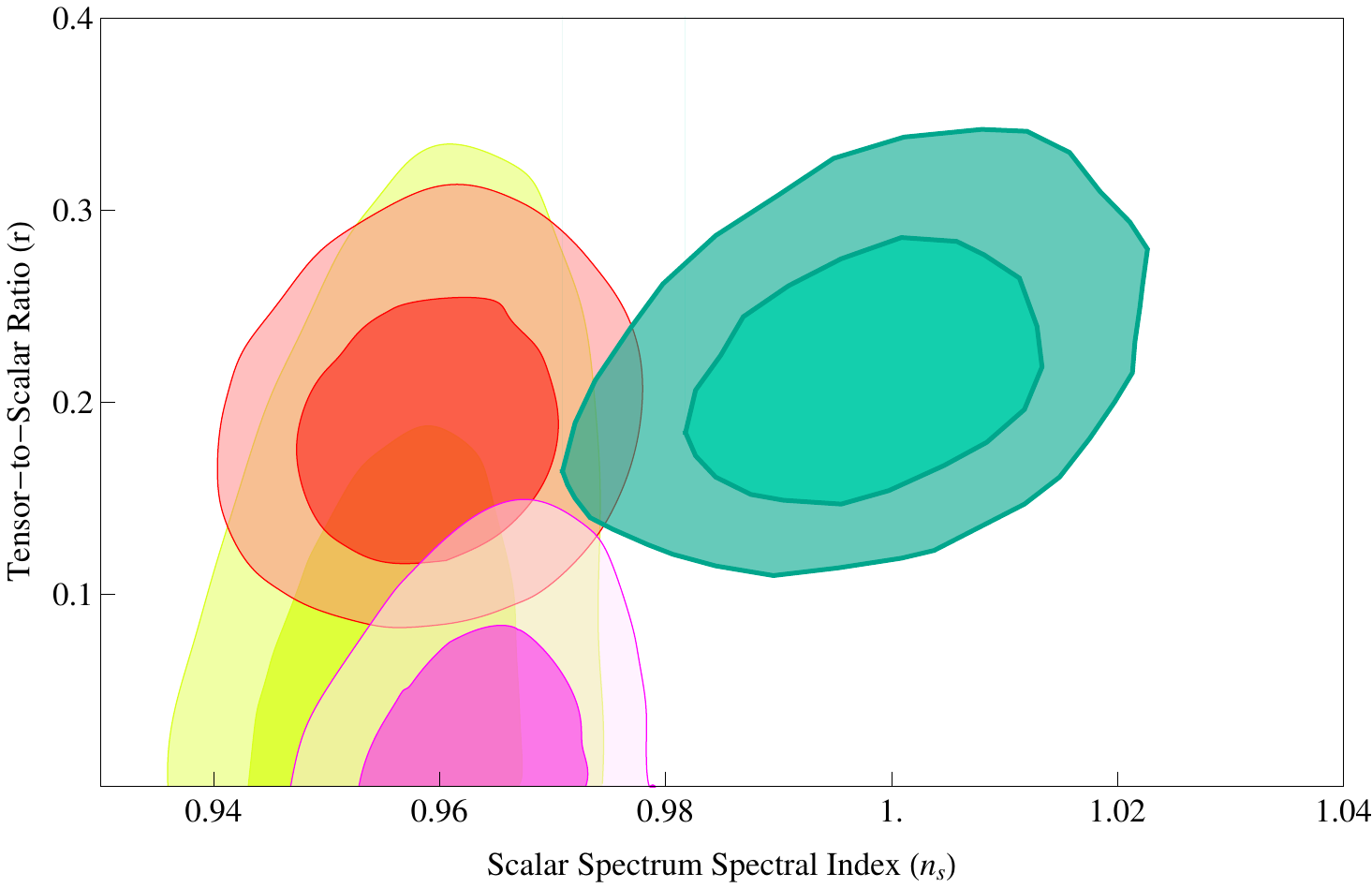}{0.99}
\end{minipage}
\begin{minipage}[t]{0.32\textwidth}
\postscript{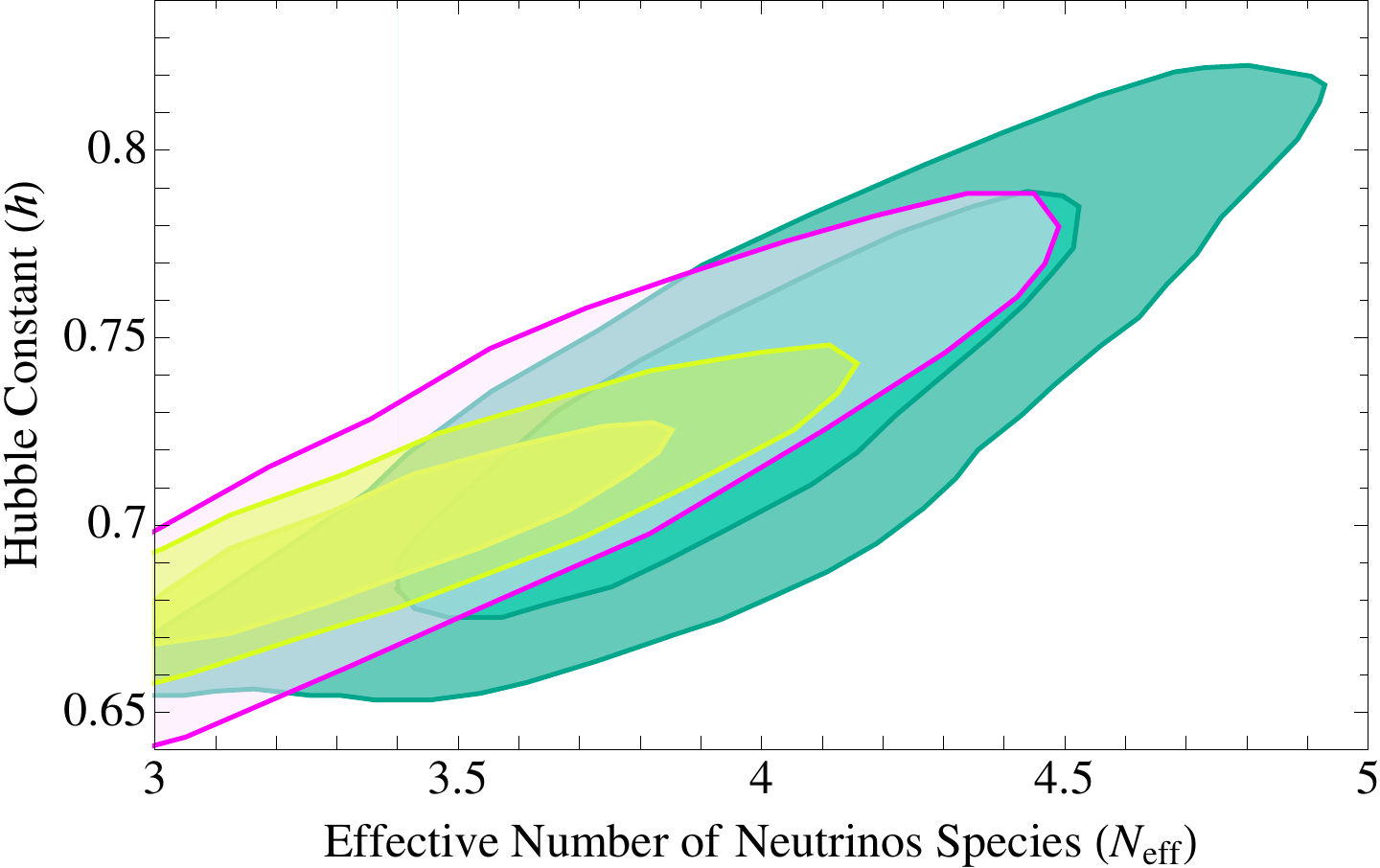}{0.99}
\end{minipage}
\begin{minipage}[t]{0.32\textwidth}
\postscript{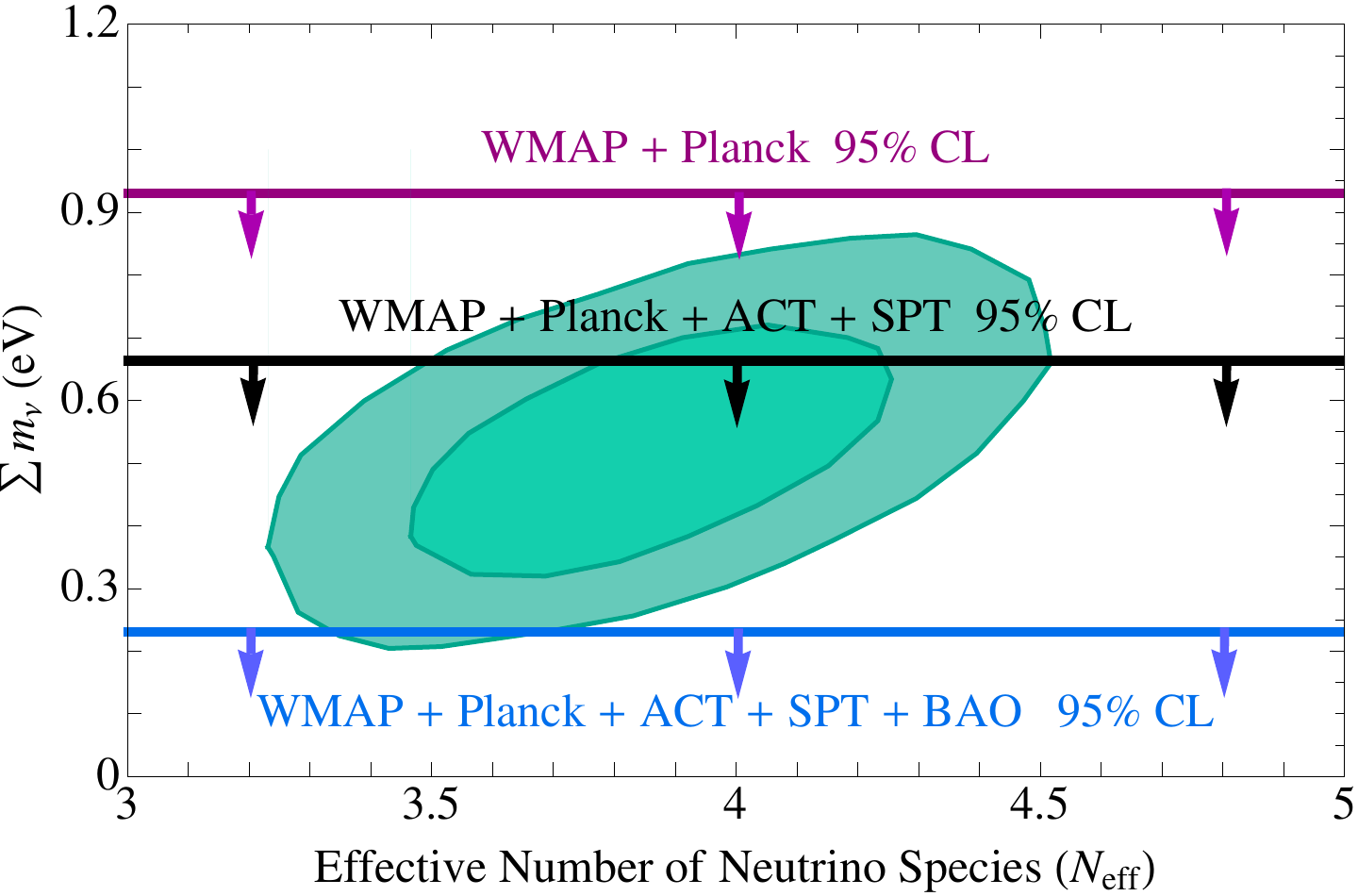}{0.99}
\end{minipage}
\caption{{\it Left:} Marginalized joint 68\% CL and 95\% CL regions
  for $(r,n_s)$ using Planck + WMAP + BAO with and without a running 
spectral index, BICEP2 data with $\alpha_s \neq
  0$ and allowed regions
  of the 9-parameter fit. {\it Middle:} 68\% and 95\%
  confidence regions for $\Lambda$CDM + $N_{\rm eff}$, using Planck +
  WMAP (pink) and Planck + WMAP + BAO (yellow) data, together with allowed
  regions of the 9-parameter fit (green) together . {\it Right:} 68\%
  and 95\% confidence regions of the 9-parameter fit. The horizontal
  lines indicate the 95\% CL upper limits on $\sum m_\nu$.}
\label{fig:dos} 
\end{figure}
\begin{figure}[tbp]
\begin{minipage}[t]{0.49\textwidth}
\postscript{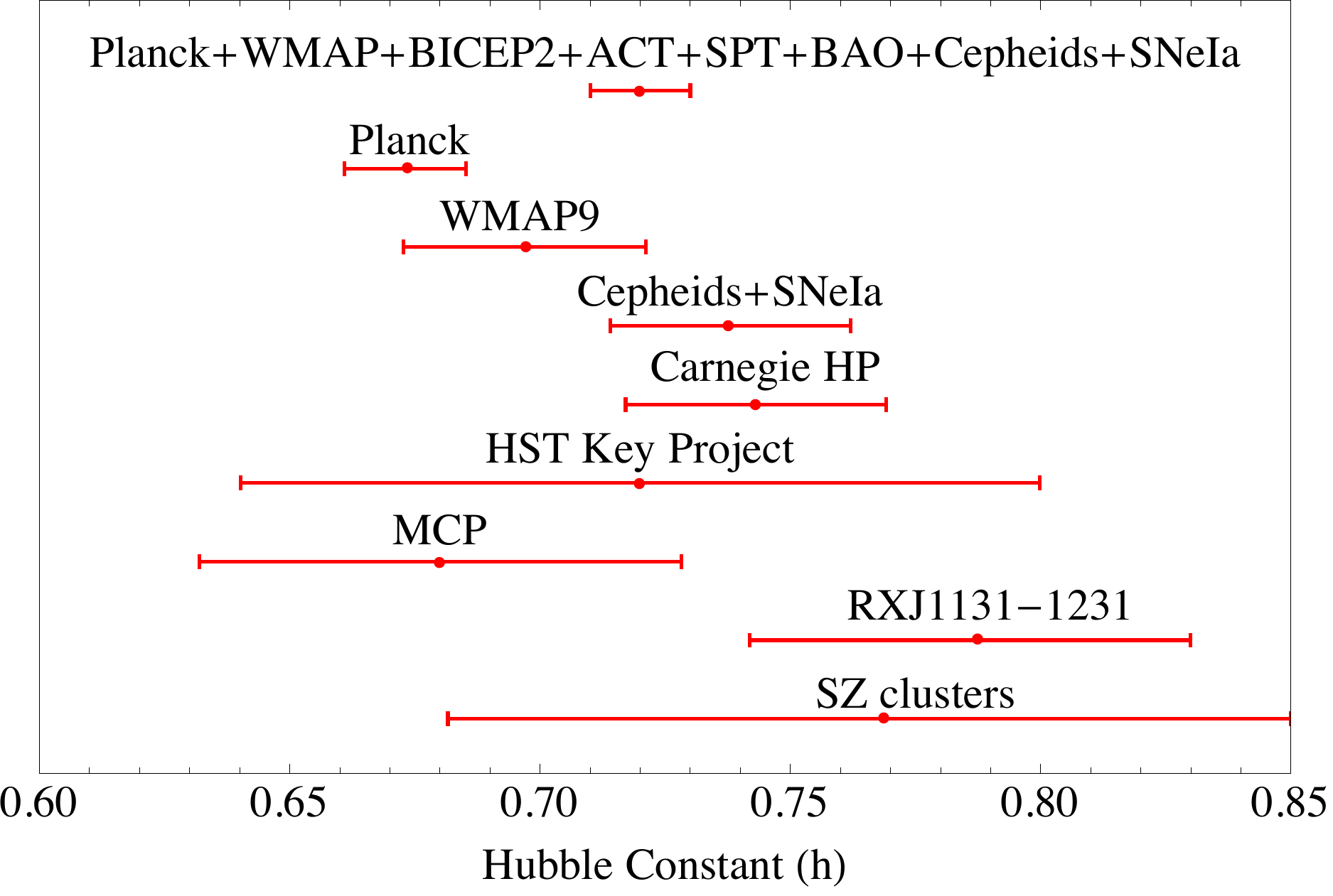}{0.99}
\end{minipage}
\hfill
\begin{minipage}[t]{0.49\textwidth}
\postscript{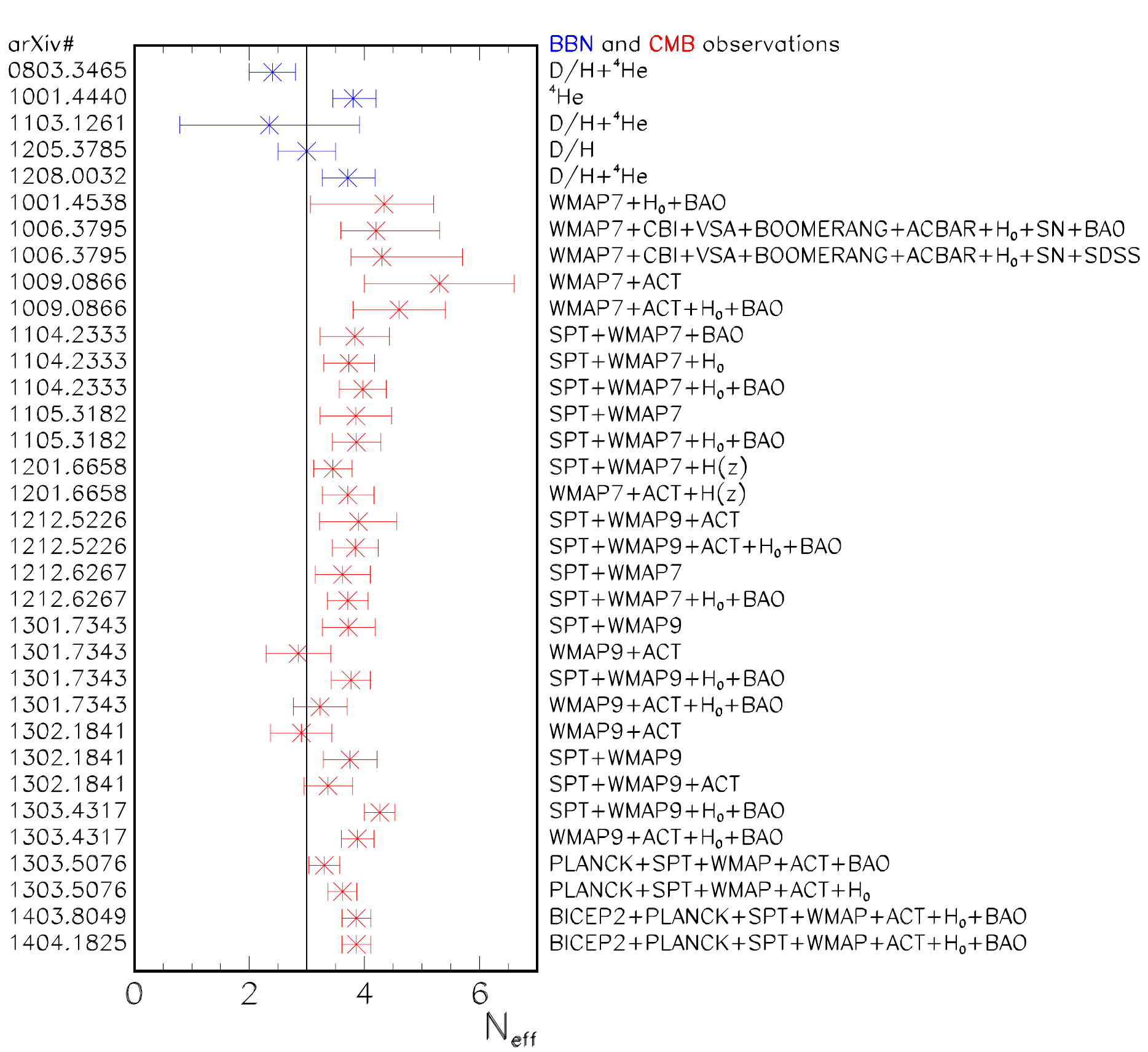}{0.99}
\end{minipage}
\caption{Recent  $H_0$ (left) and
  $N_{\rm eff}$ (right) measurements and the $1\sigma$ confidence intervals from various combinations of models and data sets.}
\label{fig:tres}
\end{figure}

As the BICEP2 Collaboration carefully emphasized (Ade et al. 2014),
the measurement of $r = 0.2^{+0.07}_{-0.05}$ (or $r =
0.16^{+0.06}_{-0.05}$ after foreground subtraction, with $r = 0$
disfavored at 5.9$\sigma$) from the B-mode polarization appears to be
in tension with the 95\% CL upper limits reported by the WMAP ($r <
0.13$, Hinshaw et al. 2009) and Planck ($r < 0.11$, Ade et al. 2013a)
collaborations from the large-scale CMB temperature power spectrum. As
shown in Fig.~\ref{fig:uno}, extension of the 7-parameter model to
include non-zero running of the spectral index ameliorates the
tension. However, the combination of Planck and BICEP2 data favors
$\alpha_s < 0$ at almost the $3\sigma$ level, with best fit value
around \mbox{$\alpha_s = -0.028 \pm 0.009$} (68\%CL) (Ade et
al. 2014). This is about 100 times larger than single-field ($\phi$) inflation
would predict. Such a particular running can be accommodated, however,
if $V '''/V$ is roughly 100 times larger than the natural expectation
from the size of $V '/V \sim (10M_{\rm Pl})^{-1}$ and $V''/V \sim (10
M_{\rm Pl})^{-2}$, where $V(\phi)$ is the inflaton potential
(Smith et al. 2014). In Fig.~\ref{fig:dos} we compare the aftermath of the
multiparameter fit of $\{\Omega_b h^2,\, \Omega_{\rm CDB} h^2,\,
\Theta_s,\, \tau,\, n_s,\, A_s,\, r,\, N_{\rm eff},\, \sum m_\nu\}$
to the data
reported by the Planck and BICEP2 collaborations (Dvorkin et al. 2014; Anchordoqui et al. 2014a). Clearly, a higher
effective number of relativistic species can relieve the tension
between Planck and BICEP2 results. As shown in Fig.~\ref{fig:tres}, the best
multiparameter fit yields $N_{\rm eff} = 0.81 \pm 0.25$ and $h = 0.70
\pm 0.01$, which are consistent with previous measurements.

We end with an observation: that one should keep in mind
that there is an on going controversy concerning the effect of
background on the BICEP2 result (Liu et al. 2014; Flauger et
al. 2014). In the next section we play devil's advocate
and assume that the BICEP2 results are flawed.

\section{$S$-dual Inflation}

\begin{figure}[tbp]
\begin{minipage}[t]{0.49\textwidth}
\postscript{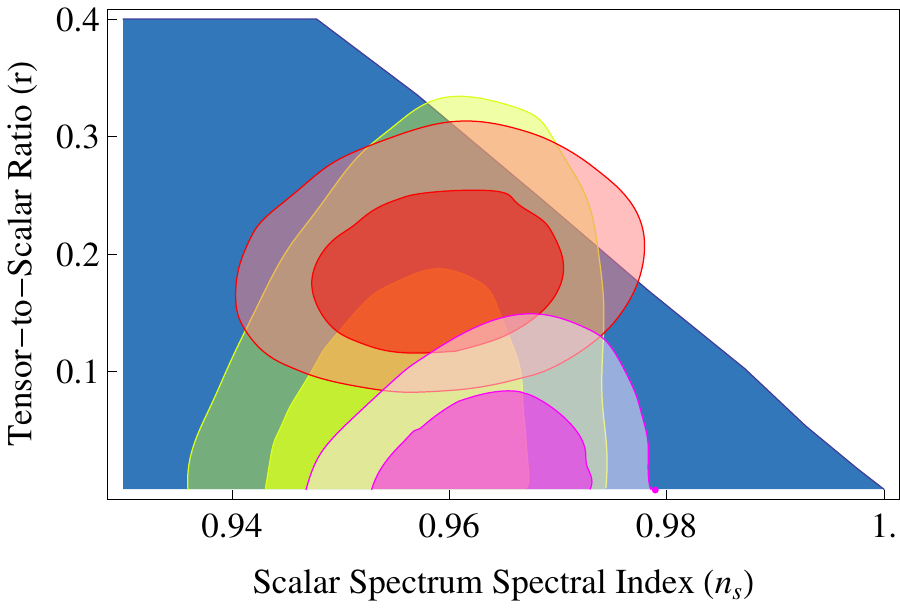}{0.99}
\end{minipage}
\hfill
\begin{minipage}[t]{0.49\textwidth}
\postscript{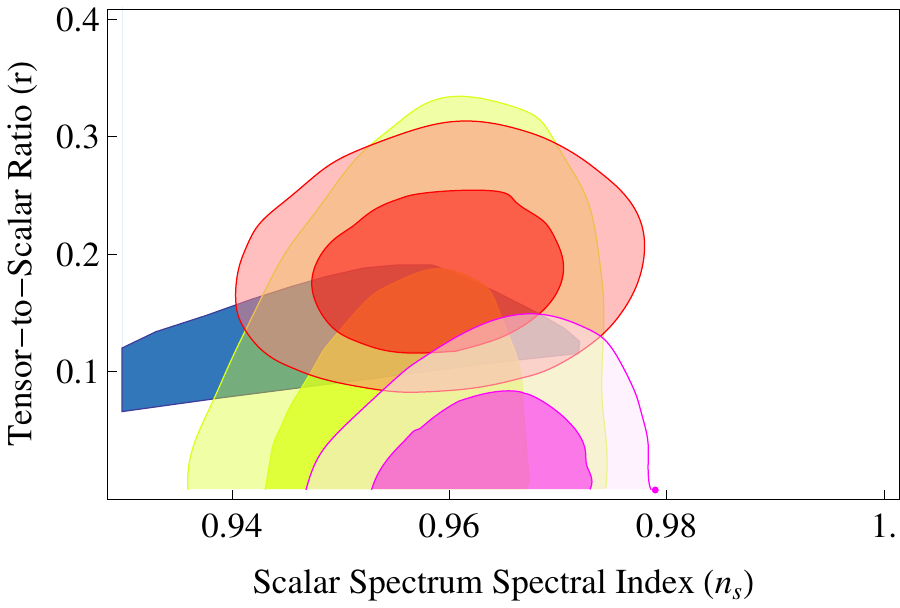}{0.99}
\end{minipage}
\caption{Available parameter space to the potential $V_1$ (left) and $V_2$ (right) together with  favored regions by Planck and BICEP2 data (Anchordoqui et al. 2014b). For $V_2$, $N>60$ corresponds to $r \lesssim 0.1$}
\label{fig:cuatro}
\end{figure}

\begin{figure}[tbp]
\begin{minipage}[t]{0.49\textwidth}
\postscript{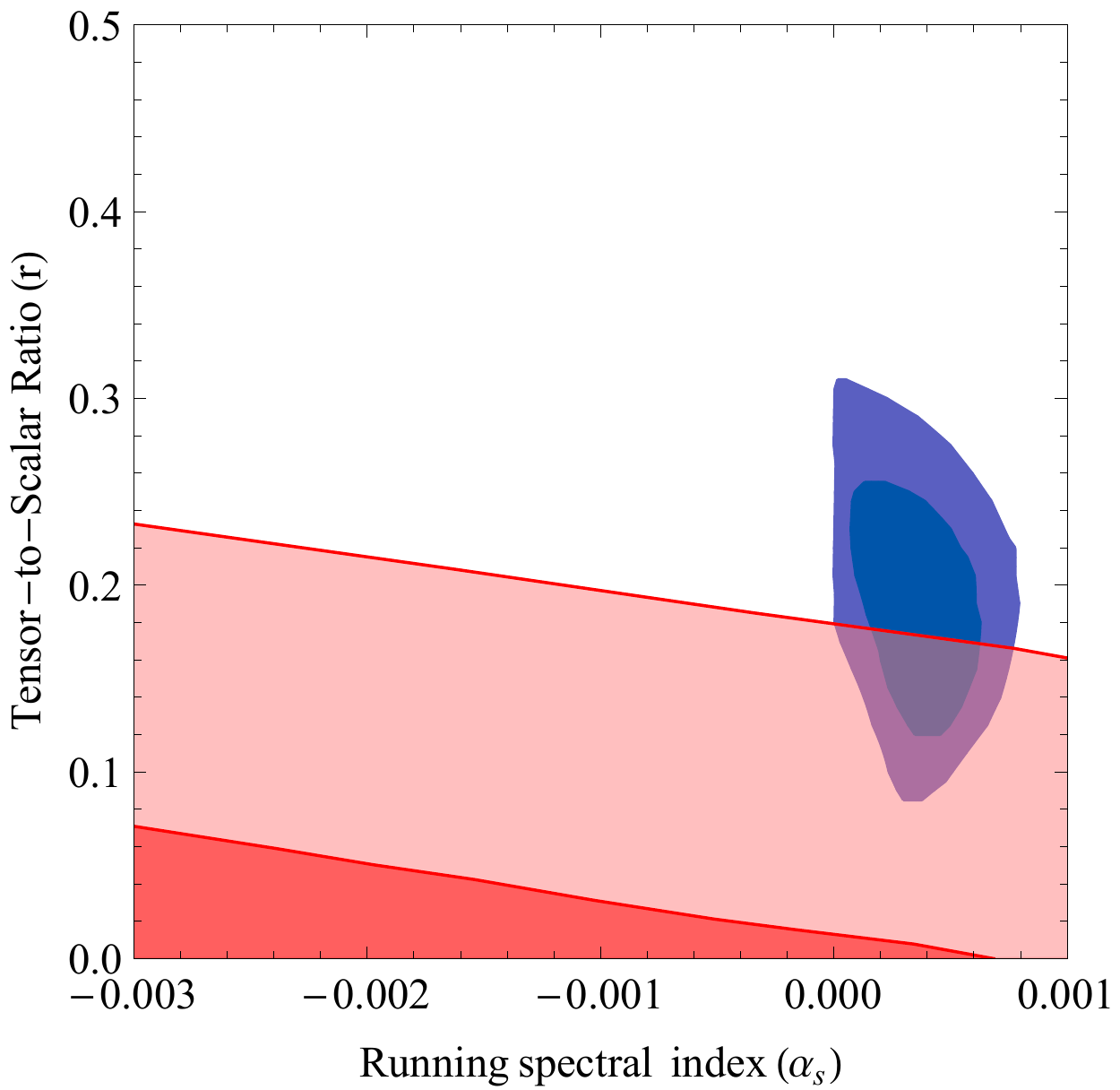}{0.99}
\end{minipage}
\hfill
\begin{minipage}[t]{0.49\textwidth}
\postscript{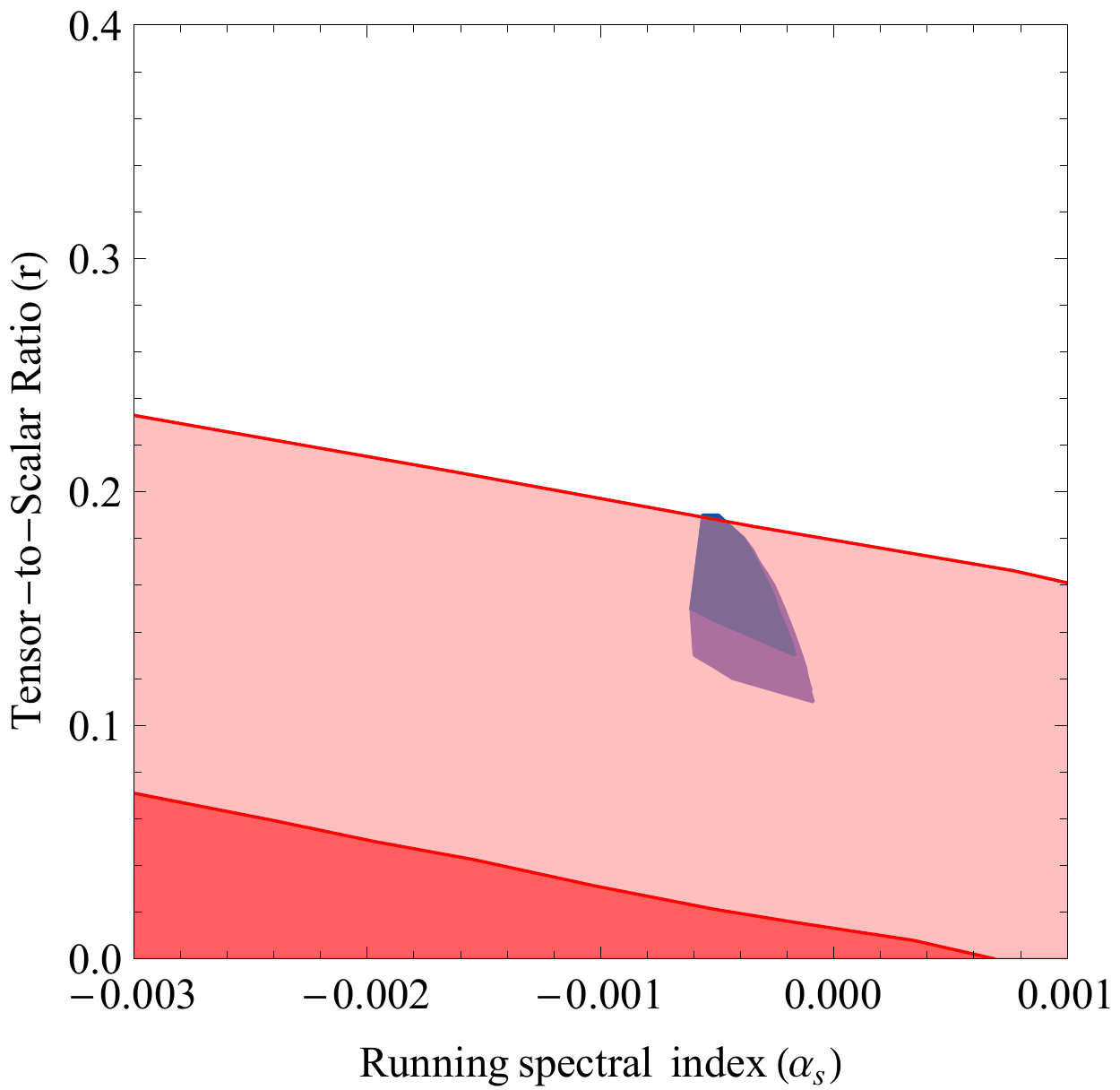}{0.99}
\end{minipage}
\caption{The blue regions show the predictions of $\alpha_2$ for the
  parameter space available to $V_1$ (left) and $V_2$ (right) after
  fixing $r$ to be within the 68\% and 95\% CL regions of BICEP2
  measurements. The red areas show the 68\% and 95\% CL regions
  favored by a combination of Planck data, WMAP polarization data and
  small scale CMB data (Ade et al. 2013a).}
\label{fig:cinco}
\end{figure}

Planck data favor standard slow-roll single field inflationary models
with plateau-like potentials $V(\phi)$ for which $V'' <0$, over
power-law potentials. However, most of these plateau-like inflaton
potentials experience the so-called ``unlikeliness problem'' (Ijjas et
al.  2013). The requirement that $V'' < 0$ in the de Sitter region,
and the avoidance of the unlikeliness problem, must now also
accommodate (if possible) the tensor-to-scalar ratio detected by
BICEP2 data.  Finally, a wish rather than a constraint: that the
inflaton potential possess some connection to particle physics. To
this end, we hypothesize that the potential be invariant under the
$S$-duality constraint $g\ra 1/g$, or $\phi\ra -\phi$, where $\phi$ is
the dilaton/inflaton, and $g\sim e^{\phi/M}$.\footnote{String theory
  exhibits various forms of dualities, i.e. relation between different
  theories at large and small radii of the compactified manifold
  (traget space duality, or $T$ duality, Giveon et al. 1994) and at
  strong and weak coupling ($S$ duality, Font et al. 1990). At the
  classical level, these dualities appear in equations of motion and
  in their solutions. Herein we do not attempt a full association with
  a particular string vacuum, but simply regard the self-dual
  constraint as a relic of string physics in big bang cosmology.} Here
$M$ is expected to be within a few orders of magnitude of $M_{\rm
  Pl}$. This requirement forces the functional form $V(\phi) = f[
\cosh(\phi/M)] $ on the potential. In what follows we take for $V$ the
$S$ self-dual form $V_1 = V_0 \, {\rm sech} (\phi/M)$, and $ V_2=V_0\
\lsb {\rm sech}(3\phi/M) - \frac 1 4 {\rm sech}^2(\phi/M)\rsb,$ which
solve the unlikeliness problem because they have no power-law
wall. For $V_1$, as for power-law inflation (with an exponential
potential), inflation does not end. We assume that the dynamics of a
second field leads to exit from the inflationary phase into the
reheating phase. The requirement that there be 50 to 60 $e$-folds of
observable inflation yields $M \gtrsim 1.4 M_{\rm Pl}$, constraining
the available region in the $r - n_s$ plane. As can be seen in
Fig.~\ref{fig:cuatro}, the allowed region is consistent with both
Planck and BICEP2 data. (Details of the calculation are given in
Anchordoqui et al. 2014b). However, as anticipated above, the
prediction for $\alpha_s$ is about 100 times smaller than the observed
68\% confidence regions, see Fig.~\ref{fig:cinco}. For $\alpha_s \neq
0$, agreement with data is only attained at 95\% CL.

\acknowledgments I thank my collaborators Vernon Barger, Haim
Goldberg, Xing Huang, Danny Marfatia,  and Brian Vlcek
for their contributions to the work discussed here. This work has been
supported by the U.S. NSF CAREER Award PHY-1053663 and by NASA Grant
No. NNX13AH52G.


\begin{references}

\reference Ade, P.A.R., et al. [BICEP2 Collaboration] 2014,
PhRvL, {\bf 112}, 241101.


\reference Ade, P.A.R., et al. [Planck Collaboration] 2013a, arXiv:1303.5082.

\reference Ade, P.A.R., et al [Planck Collaboration] 2013b, arXiv:1303.5076.

\reference Anchordoqui, L.A.,  Goldberg, H., Huang, X., \& Vlcek B.J.
2014a, JCAP, {\bf 1406}, 042.

\reference Anchordoqui, L.A., Barger, V., Goldberg, H., Huang, X., \&
Marfatia, D. 2014b, PhLB, {\bf 734}, 134.


\reference Baumann, D. 2009, arXiv:0907.5424.

\reference Dvorkin, C., Wyman, M., Rudd, D.H., \& Hu, W.
2014, arXiv:1403.8049

\reference Flauger, R., Hill, J.C., \& Spergel, D.N. 2014, arXiv:1405.7351.

\reference Font, A., Ib\'a\~nez, L.E., L\"ust, D., \& Quevedo, F. 1990,
PhLB, {\bf 249}, 35.


\reference Giveon, A., Porrati, M., \& Rabinovici, E. 1994, PhR,
{\bf 244}, 77.


\reference Hinshaw, G., et al. [WMAP Collaboration] 2013, ApJS,
{\bf 208}, 19.

\reference Ijjas, A., Steinhardt, P.J., \& Loeb, A. 2013, PhLB, {\bf 723}, 261.


\reference Liu, H., Mertsch, P., \& Sarkar, S. 2014, ApJ, {\bf 789}, L29.

\reference Riess, A.G., et al. 2011, ApJ, {\bf 730}, 119. [Erratum-ibid.\  {\bf 732}, 129, (2011)].

\reference Smith, K.M., Dvorkin, C., Boyle, L., Turok, N., Halpern, M., Hinshaw, G., \& Gold, B. 2014, PhRvL, {\bf 113}, 031301.

\reference Steigman, G., Schramm, D.N., \& Gunn, J.E. 1977, PhLB {\bf 66}, 202.

\end{references}
\end{document}